\documentstyle[12pt,epsf]{article}

\begin{document}

\begin{titlepage}

\begin{flushright}
CERN-TH/97-50\\
NORDITA 97/17-P\\
March 1997\\
hep-ph/9703389
\end{flushright}

\vspace{1cm}
\begin{center}
\Large\bf
Large-order Behaviour due to Ultraviolet Renormalons 
in QCD
\end{center}

\vspace{1cm}
\begin{center}
{\sc M. Beneke}\\[0.2cm]
{\sl Theory Division, CERN\\ 
CH-1211 Geneva 23, Switzerland}\\[0.6cm]
{\sc V.M.~Braun}\\[0.2cm]
{\sl NORDITA\\
Blegdamsvej 17, DK--2100 Copenhagen, 
Denmark}\\[0.6cm]
and\\[0.2cm]
\bigskip
{\sc N. Kivel}\\[0.2cm]
{\sl St. Petersburg Nuclear Physics Institute\\
188350 Gatchina, Russia}\\[0.3cm]
\end{center}
\vspace*{1cm}
\centerline{\bf Abstract}
\vspace*{0.2cm}
\noindent 
Ultraviolet renormalons, contrary to their infrared counterparts, 
lead to a universal contribution to the large-order 
behaviour of perturbative expansions in QCD. In this letter, 
we determine nature of the leading ultraviolet renormalon 
singularity for the inclusive hadroproduction cross section in 
$e^+ e^-$ annihilation, for hadronic $\tau$ decays and the moments 
of deep-inelastic scattering structure functions. We comment on the 
relevance of ultraviolet renormalons to estimates in low orders of 
perturbation theory.

\end{titlepage}

{\bf 1.} Perturbative expansions of physical quantities in QCD are 
divergent, and as such they are related to a measurement only through 
the additional assumption that the series is asymptotic. Of the two 
known mechanisms that cause divergence of the series, instantons 
\cite{LIP77,BOG77} are unimportant in large orders and the asymptotic 
behaviour of the series expansion is determined by renormalons 
\cite{LAU77,THO77}. According to whether the divergence arises from 
regions of small or large momentum in internal integrations, renormalons 
are classified as infrared (IR) or ultraviolet (UV). The growth of 
perturbative coefficients due to IR renormalons depends on whether 
one considers, for example, deep-inelastic structure functions, hadronic 
event shape observables or the hadronic total cross section in $e^+ e^-$
annihilation and is closely related to power corrections to these 
observables. For this reason they have recently been studied intensely  
\cite{revs}. UV renormalons are often regarded to be more 
complicated and have mostly been left aside, perhaps because they are 
Borel summable. However, the structure of UV renormalons is 
universal in the sense that it depends only on the theory under 
consideration, in our case QCD. Moreover, for some quantities of interest, 
such as the hadronic total cross section in $e^+ e^-$
annihilation, the sign-alternating UV renormalon behaviour determines 
the actual large-order behaviour of the series expansion. In this 
letter we determine the universal ultraviolet renormalon 
asymptotics in QCD and apply it to a number of observables of 
phenomenological interest.

The universality of UV renormalons was recognised by Parisi \cite{PAR78}, 
who noted that UV renormalons could be compensated by adding 
higher-dimension operators to the Lagrangian just 
as logarithmic UV divergences can be compensated by dimension-four 
counterterms. To state Parisi's hypothesis precisely, we consider 
a quantity $R(\alpha_s)$, expanded as 
\begin{equation}
\label{expansion}
R(\alpha_s) = A \left(1+\sum_{n=0}^\infty r_n\alpha_s^{n+1}\right).
\end{equation}
The dependence on external momenta is not indicated and 
$\alpha_s$ denotes the coupling renormalised at a scale $\mu$. The 
UV renormalons produce poles at $t=m/\beta_0<0$ in the 
Borel transform $B[R](t)=\sum_{n=0} r_n t^n/n!$, where $m$ is a positive 
integer and $\beta_0=-b/(4\pi)$ ($b=11-2N_f/3$) the first coefficient 
of the $\beta$-function. It follows that the 
integral\footnote{Since UV renormalons produce sign-alternating 
factorial behaviour, they do not lead to ambiguities in the usual 
Borel integral. Because the consideration of ambiguities will 
simplify the renormalisation group considerations below, we define 
the integral parallel to the negative axis. The 
imaginary parts created by UV renormalon poles are now 
exponentially large in $\alpha_s$. This need not bother us, because 
$\alpha_s$ could be considered negative without any change in our 
derivation.} 
\begin{equation}
\label{irr}
I[R](\alpha_s) = \int\limits_{0+i\epsilon}^{-t_c+i\epsilon} d t\,
e^{-t/\alpha_s}\,B[R](t) \qquad t_c>0
\end{equation}
is complex, for $t>-1/\beta_0$, 
and its imaginary part is unambiguously related to UV 
renormalon singularities, and therefore asymptotic behaviour, 
and vice versa.
In the following, we will be concerned explicitly only with the 
leading UV renormalon singularity at $t=1/\beta_0$. The hypothesis of 
Parisi \cite{PAR78}, for this situation, states that
\begin{equation}
\label{parisi}
\mbox{Im}\,I[R](\alpha_s,p_k) = \frac{1}{\mu^2}\sum_i
e^{-1/(\beta_0\alpha_s))}\alpha_s^{-\beta_1/\beta_0^2}\,
C_i(\alpha_s)\,R_{{\cal O}_i}(\alpha_s,p_k).
\end{equation} 
Thus, the leading UV renormalon behaviour is determined by 
single zero-momentum insertions of dimension-six operators\footnote{
The fact that higher (than four) dimensional operators start at 
dimension six determines that the position of the leading UV singularity  
is indeed at $t=1/\beta_0$.} ${\cal O}_i$ 
into the Green function from which 
the quantity $R$ is derived. The important point here is that 
the coefficient function $C_i$ is universal -- it is independent 
of the external momenta $p_k$ and the quantity $R$ -- which is related 
to the fact that UV renormalons arise from loop momentum regions 
much larger than any external scale. The dimension-six operators 
may be thought of as an additional term 
\begin{equation}
\Delta {\cal L} = -\frac{i}{\mu^2}\sum_i
e^{-1/(\beta_0\alpha_s))}\alpha_s^{-\beta_1/\beta_0^2}\,
C_i(\alpha_s)\,{\cal O}_i
\end{equation}
in the QCD Lagrangian with coefficients such that for any $R$ the 
imaginary part of $I[R]$ is compensated by the additional 
contribution to $R$ from $\Delta {\cal L}$. 
By comparing the renormalisation 
group equations for $I[R]$ and $R_{{\cal O}_i}$ one derives that 
\begin{equation}
\label{rge}
\frac{d}{d\alpha_s}\,C_j(\alpha_s) = \frac{\gamma_{ij}(\alpha_s)}
{2\beta(\alpha_s)}\,C_i(\alpha_s),
\end{equation}
where $\gamma(\alpha_s)$ is the anomalous dimension matrix 
pertaining to the dimension-six operators ${\cal O}_i$. The 
solution to this equation determines completely the 
$\alpha_s$-dependence of the $C_i$. This 
in turn allows us to compute the Borel transform $B[R](t)$ 
in the vicinity of the singular point $t=1/\beta_0$ up 
to an over-all constant. Finally, the nature of the singularity 
determines the large-order behaviour. These manipulations can be summarised 
in the substitution rule 
\begin{equation}
\label{subst}
e^{-1/(\beta_0\alpha_s)}\,(\beta_0\alpha_s)^\lambda\longrightarrow
\frac{1}{\pi}\sum_n\beta_0^n\,n!\,n^{-\lambda}\,\alpha_s^{n+1}
\end{equation}
to obtain the leading asymptotic behaviour of $R$ 
from the $\alpha_s$-dependence 
of $\mbox{Im}\,I[R]$.

The leading UV renormalon divergence has been subjected to 
detailed diagrammatic study in abelian models \cite{VAI94}--\cite{BEN96}, 
which confirmed the general structure of (\ref{parisi}). 
In \cite{VAI94} four-fermion operators were identified as sources 
of leading behaviour in QED. In \cite{BEN96} (\ref{parisi}) was shown 
to be valid in QED to all orders in the $1/N_f$ expansion and it was 
indicated how one could understand the restoration of the full 
non-abelian $\beta_0$ starting from a $1/N_f$ expansion of the 
non-abelian theory. Both \cite{VAI94,BEN96} corroborated the 
earlier expectation \cite{GRU93}--\cite{BEV93} that while (\ref{rge}) 
determines the $\alpha_s$-dependence of $C_i$, the integration 
constants, which are related to the constants that specify the 
over-all normalisations $K_i$ of the UV renormalon asymptotic behaviour, 
could not be calculated systematically, except in expansions 
like $1/N_f$. Once this is realized, all exact obtainable information 
can be obtained solely by solving the renormalisation group equations 
for $C_i$ above. This problem has already been solved for 
Heavy Quark Effective Theory \cite{BEN95a} and QED \cite{VAI94,BEN96}.  
In the following sections we treat the slightly more complicated 
case of QCD. That is, for 
\begin{equation}
\label{as}
r_n \stackrel{n\to \infty}{=} \beta_0^n\,n!\,n^{\beta_1/\beta_0^2}
\sum_i K_i\,n^{\delta_i}\left(1+O(1/n)\right)
\end{equation}
we determine the so far unknown constants $\delta_i$. Note that 
the $1/n$ corrections to the asymptotic behaviour are in principle 
calculable as well. This would require the two-loop anomalous 
dimension matrix as well as the one-loop corrections to the 
Green functions with operator insertion,  
$R_{{\cal O}_i}$. The constants $K_i$ remain 
unknown. However, because of the universality of coefficients 
functions $C_i$, the ratio of $K_i$'s for different observables 
is calculable.\\

{\bf 2.} In this section we determine the leading UV renormalon 
behaviour of current correlation functions 
\begin{equation}
i\int d^4 x\,e^{iq x}\,\langle 0|T(j_\mu(x) j_\nu(0))|0\rangle 
=(q_\mu q_\nu-g_{\mu\nu} q^2) \, \Pi(q^2)
\end{equation}
in massless QCD with $N_f$ flavours. We will consider colour-singlet 
vector and axial-vector currents and let them also 
be flavour-singlets.  In the following section, we generalise to the 
real case where flavour symmetry is broken by electric or axial 
charges or the 
current is flavour non-diagonal. Thus, 
in expressions like $(\bar{\psi} M\psi)$ a sum over flavour, colour and 
spinor indices is implied and $M$ is a matrix in colour and spinor space, 
but unity in flavour space.

To account for the external currents, we introduce two $U(1)$ background 
fields that couple to the vector and axial-vector current and consider 
the Lagrangian
\begin{equation}
{\cal L}_{QCD} + j_V^\mu v_\mu + j_A^\mu a_\mu.
\end{equation}
The corresponding field strength tensors are defined as usual and 
denoted by  
$F_{\mu\nu}$ and $H_{\mu\nu}$, respectively, such that 
$\partial_\mu F^{\mu\nu}=j_V^\nu$ and 
$\partial_\mu H^{\mu\nu}=j_A^\nu$. A basis of dimension-six 
operators is then given by
\begin{eqnarray}
\label{basis}
{\cal O}_1 &=& (\bar{\psi}\gamma_\mu\psi) (\bar{\psi}\gamma^\mu\psi)
\nonumber\\
{\cal O}_2 &=& (\bar{\psi}\gamma_\mu\gamma_5\psi) (\bar{\psi}
\gamma^\mu\gamma_5\psi)
\nonumber\\
{\cal O}_3 &=& (\bar{\psi}\gamma_\mu T^A\psi) (\bar{\psi}\gamma^\mu 
T^A \psi)
\\
{\cal O}_4 &=& (\bar{\psi}\gamma_\mu\gamma_5 T^A\psi) (\bar{\psi}
\gamma^\mu\gamma_5 T^A\psi)
\nonumber\\
{\cal O}_5 &=& \frac{1}{g}\,f_{ABC} \,G_{\mu\nu}^A G_\rho^{\nu\,B} 
G^{\rho\mu\,C} 
\end{eqnarray}
\vspace*{-0.7cm}
\begin{eqnarray}
{\cal O}_6 &=& \frac{1}{g^2}\,(\bar{\psi}\gamma_\mu\psi)\,
\partial_\nu F^{\nu\mu}
\qquad
{\cal O}_7 \,=\, \frac{1}{g^2}\,(\bar{\psi}\gamma_\mu\gamma_5\psi)\,
\partial_\nu H^{\nu\mu}
\\
{\cal O}_8 &=& \frac{1}{g^4}\,\partial_\nu F^{\nu\mu}\,
\partial^\rho F_{\rho\mu}
\qquad\quad\!\!
{\cal O}_9 \,=\, \frac{1}{g^4}\,\partial_\nu H^{\nu\mu}\,
\partial^\rho H_{\rho\mu}.
\end{eqnarray}
We need not consider gauge-variant operators and those that vanish by 
the equations of motion as their zero-momentum insertions into the 
above correlation functions do not contribute. The over-all factors 
$1/g^k$ have been inserted for convenience. Note that we did not 
include four-fermion operators of scalar, pseudo-scalar or tensor 
type. They can not be generated in massless QCD, because 
the number of Dirac matrices on any fermion line that connects 
to an external fermion in a four-point function is always odd. 
The coefficients $C_i$ corresponding to these operators therefore 
vanish exactly. It is straightforward to 
compute the leading-order anomalous dimension matrix $\gamma$, defined such 
that the renormalised operators satisfy
\begin{equation}
\left(\delta_{ij}\,\mu\frac{d}{d\mu}+\gamma_{ij}\right)
{\cal O}_j = 0.
\end{equation}
We find $\gamma=\gamma^{(1)}\alpha_s/(4\pi)$ with\footnote{The 
renormalisation of dimension-six operators has been studied previously in 
various contexts, mainly QCD sum rules. We recalculated the entries 
of $\gamma^{(1)}$ in our basis, except for $\gamma_{55}$, which we 
take from \cite{GGG}.}
\begin{equation}
\label{matrix}
\gamma^{(1)} = \left(
\begin{array}{ccc}
A & 0 &B \\
0 & \gamma_{55} & 0\\
0 & 0 & C
\end{array}\right) 
\end{equation}
and ($C_F=(N_c^2-1)/(2 N_c)$, $N_c$ the number of colours)
\begin{equation}
A=\left(
\begin{array}{cccc}
0 & 0 & \frac{8}{3} & 12 \\[0.1cm]
0 & 0 & \frac{44}{3} & 0 \\[0.1cm]
0 & \frac{6 C_F}{N_c} & -\frac{9 N_c^2+4}{3 N_c}+\frac{8 N_f}{3} & 
\frac{3 (N_c^2-4)}{N_c} \\[0.1cm]
\frac{6 C_F}{N_c} & 0 &\frac{3 (N_c^2-4)}{N_c} -\frac{4}{3 N_c} & 
-3 N_c
\end{array}\right) .
\end{equation} 
The non-zero entries of the $4\times4$ sub-matrices $B,C$ are: 
$B_{11}=B_{22}=8 (2 N_c N_f+1)/3$, $B_{12}=B_{21}=8/3$, 
$B_{31}=B_{32}=B_{41}=B_{42}=8 C_F/3$, $C_{11}=C_{22}=-2 b$, 
$C_{33}=C_{44}=-4 b$, $C_{13}=C_{24}=8 N_c N_f/3$. The mixing of 
${\cal O}_5$ into itself is given by $\gamma_{55}=-8(N_c-N_f)/3$ 
\cite{GGG}. Note that ${\cal O}_5$ could mix into ${\cal O}_3$ at 
order $\alpha_s$. We find that due to a cancellation of different 
diagrams the corresponding entry $\gamma_{53}$ vanishes. As a 
consequence, ${\cal O}_5$ decouples from the mixing at leading order, 
in agreement with the first reference of \cite{GGG}.

To solve (\ref{rge}) with $\gamma(\alpha_s)$ and $\beta(\alpha_s)$ 
evaluated at leading order let $2 b\lambda_i$, 
$i=1\ldots 4$, be the eigenvalues of $A$ and $\lambda_5=\gamma_{55}/(2 b)$. 
Let $U$ be the matrix that diagonalises $A$. Since the integration 
constants must be considered as non-perturbative, we need not 
keep track of factors multiplying these constants, unless they 
are exactly zero. Thus we only note that no element of $U$ vanishes 
for values of $N_f$ of interest. Since $C$, and therefore 
$\gamma^{(1)}$, is triangular, we readily obtain
\begin{eqnarray}
C_i(\alpha_s) &=& \sum_{k=1}^4 C^{[1]}_{ik}\alpha_s^{-\lambda_k}
\qquad i=1\ldots 4
\nonumber\\
C_5(\alpha_s) &=& C^{[1]}_5\alpha_s^{-\lambda_5} 
\\
C_i(\alpha_s) &=& C^{[2]}_i\alpha_s+ 
\sum_{k=1}^4 C^{[1]}_{ik}\alpha_s^{-\lambda_k}
\qquad i=6,7
\nonumber\\
C_i(\alpha_s) &=& C^{[2]}_i\alpha_s+ C^{[3]}_i\alpha_s^2 + 
\sum_{k=1}^4 C^{[1]}_{ik}\alpha_s^{-\lambda_k}
\qquad i=8,9
\nonumber
\end{eqnarray}
with $\alpha_s$-independent non-vanishing constants $C^{[l]}$ that depend 
on nine integration constants and the elements of $\gamma^{(1)}$. The 
exponents $\lambda_k$ are reported in Table~\ref{tab1}. The values 
for $\lambda_1$ to $\lambda_4$ at $N_f=3$ are in agreement with 
\cite{svz}. We emphasise again 
that the coefficient functions $C_i$ are independent of the particular 
Green function or observable under consideration.

\begin{table}[t]
\addtolength{\arraycolsep}{0.2cm}
\renewcommand{\arraystretch}{1.25}
$$
\begin{array}{c|ccccc}
\hline\hline
N_f & \lambda_1 & \lambda_2 &\lambda_3 &\lambda_4 & \lambda_5  \\ 
\hline
3 & 0.379 & 0.126 & -0.332 & -0.753 & 0  \\
4 & 0.487 & 0.140 & -0.302 & -0.791 & 4/25 \\
5 & 0.630 & 0.155 & -0.275 & -0.843 & 8/23 \\
6 & 0.817 & 0.172 & -0.254 & -0.910 & 4/7    \\ 
\hline\hline
\end{array}
$$
\caption{\label{tab1}
Numerical values of $\lambda_i$ ($N_c=3$).}
\end{table}

We now consider $R=q^2d\Pi(q^2)/dq^2$, the `Adler functions' derived from 
the correlation functions of two vector or axial-vector currents and 
write their perturbative expansions as in (\ref{expansion}). Having 
found the $\alpha_s$-dependence of the coefficient functions, we 
further require $R_{{\cal O}_i}(\alpha_s,q)$ at leading order. 
Up to constants, we have $R_{{\cal O}_i}(\alpha_s,q)\propto \alpha_s^0$, 
$i=1\ldots 4$, $R_{{\cal O}_5}(\alpha_s,q)\propto \alpha_s$, 
$R_{{\cal O}_i}(\alpha_s,q)\propto \alpha_s^{-1}$, $i=6,7$ and  
$R_{{\cal O}_i}(\alpha_s,q)\propto \alpha_s^{-2}$ for $i=8,9$. Combining 
all factors in (\ref{parisi}) and taking into account the 
substitution rule (\ref{subst}) 
to obtain the large-order behaviour of $R$ from the imaginary 
part of $I[R]$ our final result reads
\begin{equation}
\label{asfinal}
r_n \stackrel{n\to \infty}{=} \beta_0^n\,n!\,n^{\beta_1/\beta_0^2}
\left[\,\sum_{i=1}^4 K_i\,n^{2+\lambda_i} + K_5\,n^{-1+\lambda_5} +
K_6 + K_8\, n\right]\left(1+O(1/n)\right)
\end{equation}
for two vector currents.\footnote{We display in (\ref{asfinal}) the general 
structure of the result, although it is not consistent to keep 
all terms without computing $1/n$-corrections to the leading 
contribution proportional to $K_1$.}  
The result for two axial currents is identical, 
except for different over-all constants $K_i$. Note that the leading 
behaviour is generated by the largest eigenvalue of the anomalous 
dimension matrix of four-fermion operators. The contribution from the 
three-gluon operator (proportional to $K_5$) 
is suppressed. The dominant contribution from this 
operator comes in fact from its mixing into ${\cal O}_{6/7}$ at 
next-to-leading order (not computed here) rather than the `direct' 
contribution through $R_{{\cal O}_5}(\alpha_s)$ 
in (\ref{asfinal}) above. In general, such 
next-to-leading order contributions result in $1/n$-corrections to 
the asymptotic behaviour (\ref{asfinal}). Eq.~(\ref{asfinal}) holds 
when the series is expressed in terms of the $\overline{\mbox{MS}}$ 
renormalised coupling $\alpha_s$, the convention we assume throughout this 
note. If a different coupling is employed that is related 
to the $\overline{\mbox{MS}}$ coupling by a factorially divergent 
series, the coefficients $r_n$ change accordingly and 
(\ref{asfinal}) may not be valid.\\

{\bf 3.} The previous result for flavour-$\mbox{U}(N_f)$ singlet 
currents applies practically unaltered to all observables of interest. 
We briefly discuss them case by case.

\paragraph{$e^+ e^-$ annihilation into hadrons.} Consider first 
annihilation through virtual photons. The external current is 
$j_\mu=\bar{\psi}\gamma_\mu Q\psi$, where $Q_{ij}=
\mbox{diag}(e_u,e_d,\ldots)$ is a matrix in flavour space and 
flavour indices are summed over. Since flavour 
symmetry is broken only by the external current (all quarks are 
still considered as massless), the `QCD operators' ${\cal O}_{1-5}$ 
remain unaltered. The basis of `current operators' ${\cal O}_{6-9}$ 
has to be altered to include the operators $(\mbox{tr} Q)
\,\bar{\psi}\gamma^\mu
\psi\,\partial_\nu F^{\nu\mu}$ and $\bar{\psi}\gamma^\mu Q 
\psi\,\partial_\nu F^{\nu\mu}$ instead of ${\cal O}_6$. (Similar 
modifications would apply for the axial-vector current.) This ensures that 
mixing of four-fermion operators into the current operators contributes 
proportional to $\mbox{tr} Q^2=\sum_f e_f^2$ and 
$(\mbox{tr} Q)^2=(\sum_f e_f)^2$, as 
required by the existence of `flavour non-singlet' and `light-by-light 
scattering' terms. The matrices $B$ and $C$ in (\ref{matrix}) change, 
but their pattern of non-zero entries does not. Thus, as we are not  
interested in over-all constants, (\ref{asfinal}) carries over to 
the present case. The leading asymptotic behaviour of the 
Adler function, expressed as in (\ref{expansion}), is 
\begin{equation}
\label{adler}
d_n \stackrel{n\to \infty}{=} K_d\,\beta_0^n\,n!\,
n^{2+\beta_1/\beta_0^2+\lambda_1} = K_d\,\beta_0^n\,n!\,n^{1.97},
\end{equation}
where in the last line we have taken $N_f=5$. This is 
to be compared with the large-$N_f$ limit \cite{BEN93} which 
is recovered by setting $n^{1.97}\to n$. Eq.~(\ref{adler}) holds 
separately for the `flavour non-singlet' and `light-by-light 
scattering' contributions. In large orders, the expansion 
of the $e^+ e^-$ hadronic cross section is related to the Adler 
function by 
\begin{equation}
\label{r}
r_n^{e^+ e^-} \stackrel{n\to \infty}{\propto} \frac{d_n}{n}.
\end{equation}
The suppression by one power of $n$ follows from the fact that 
the coefficient of the leading asymptotic behaviour of the $d_n$ 
is polynomial in the external momentum and therefore 
does not contribute to 
the discontinuity. 

If we now consider the hadronic width of the $Z^0$, the inclusion of 
axial currents proceeds along the same lines as above for the 
vector current and modifies over-all constants, but again the general 
pattern of mixing remains the same. The expansion coefficients of the 
$Z^0$ width therefore grow as $r_n^{e^+ e^-}$ in large orders. 
Note that the fact that flavour-singlet terms arise at 
order $\alpha_s^2$ for axial currents, but $\alpha_s^3$ for 
vector currents does not affect large-order estimates. 

\paragraph{Hadronic $\tau$ decay.} For flavour non-singlet currents 
$j_\mu=\sum_{f=d,s} V_{uf}\,\bar{u}\gamma_\mu(1-\gamma_5) q_f$  
the `light-by-light scattering' diagrams are absent. For an 
appropriate basis of current operators this is reflected in a 
change of entries in the matrices $B$ and $C$. However, their pattern 
of non-zero entries is not changed. The contour integral that 
relates the $\tau$ width to the Adler function of the currents 
suppresses the large-order behaviour just as in case of the 
$e^+ e^-$ cross section above. We therefore have 
\begin{equation}
\label{tau}
r_n^\tau \stackrel{n\to \infty}{\propto} 
r_n^{e^+ e^-} \stackrel{n\to \infty}{\propto} 
\beta_0^n\,n!\,n^{0.59}.
\end{equation}
In the present case we set $N_f=3$.

\paragraph{Moments of deep-inelastic scattering (DIS) structure 
functions.} The operator product expansion allows us to write 
(at leading twist)
\begin{equation}
\int_0^1dx\,x^{N-1}\,F_k(x,Q) = \sum_i C^i_{k,N}\!\left(\frac{Q^2}{\mu^2},
\alpha_s\right)\,A_N^i(\mu).
\end{equation}
Here the structure function 
$F_k$ can be $2 F_1$, $F_2/x$ or $F_3$, $A_N^i(\mu)$ denotes 
(reduced) proton matrix elements of twist-2 operators and 
$C^i_{k,N}(\alpha_s)$ the coefficients functions. To compute the 
large-order behaviour of the perturbative expansion of coefficient 
functions, we need to compute the insertion of 
dimension-six operators into the quark matrix 
elements of current-current correlation functions.\footnote{Gluon 
matrix elements are suppressed and do not contribute to the 
leading large-order behaviour.} Because we consider quark matrix 
elements (rather than vacuum matrix elements as above), insertions 
of current-current operators such as ${\cal O}_{8/9}$ vanish and the 
leading asymptotic behaviour arises from insertion of operators 
such as ${\cal O}_{6/7}$. Furthermore, collinear singularities have 
to be factorised by computing the analogues of $A_N^i(\mu)$ between 
quark states and dividing them out. We choose the  
$\overline{\mbox{MS}}$ factorisation scheme, in which case 
the $A_N^i(\mu)$ are pure poles. Dividing them out then does not 
modify the large-order behaviour of the finite terms. Finally, 
taking moments results in a moment-dependent over-all 
constant, but the $n$-dependence is the same for all moments. 
Thus,
\begin{equation}
\label{dis}
C^i_{k,N}(\alpha_s)\stackrel{n\to \infty}{=} \sum_n K^i_{k,N}\,
\beta_0^n\,n!\,
n^{1+\beta_1/\beta_0^2+\lambda_1}\,\alpha_s^{n+1}.
\end{equation}
Compared to the Adler function in (\ref{adler}) one has one power of $n$ 
less, because the operators ${\cal O}_{8/9}$ do not contribute. 
This reflects that at a given order in $\alpha_s$ the diagrams that 
contribute to the coefficient functions have one loop less compared 
to the diagrams that contribute to the Adler function. Eq.~(\ref{dis}) 
applies in particular to the case of QCD corrections to the GLS and 
Bjorken sum rule.\\

In general, the contribution from IR renormalons to the asymptotic 
behaviour can compete with the UV renormalon asymptotics computed 
in this note. For the moments of DIS structure functions, one finds 
$r_n^{IR}/r_n^{UV}\sim (-1)^n$ and the nature of the IR renormalon 
singularity is determined by the anomalous 
dimensions of twist-4 operators \cite{MUE85}.
For the special case of the GLS sum rule, using the anomalous 
dimension calculated in  
\cite{SV82}, we obtain
\begin{equation}
\label{dis2}
C_{GLS}(\alpha_s)\stackrel{n\to \infty}{=} \sum_n 
\beta_0^n\,n!\left[K^{UV}_{GLS}\,n^{1+\beta_1/\beta_0^2+\lambda_1} + 
K^{IR}_{GLS}\,(-1)^n\,n^{-\beta_1/\beta_0^2-(4/3b)(N_c-1/N_c)}
\right]\alpha_s^{n+1}.
\end{equation}
For the interesting case of $N_f>2$, the UV renormalon behaviour formally 
dominates at very large $n$. 
For $e^+ e^-$ annihilation, $\tau$ and $Z^0$ 
decay the first IR renormalon singularity occurs at 
$t=-2/\beta_0$ \cite{PAR79} and thus its contribution to the 
asymptotic behaviour is suppressed as $r_n^{IR}/r_n^{UV}\sim 
(-2)^{-n}$. Thus, UV renormalons determine the asymptotic behaviour 
in all cases considered here. 
However, IR renormalons tend to have large over-all 
normalisation factors in the $\overline{\mbox{MS}}$ scheme as compared 
to UV renormalons \cite{BEN93a,BZ92}. Thus, dominance of IR renormalons 
at intermediate $n$ could be expected and 
is indeed observed in the fixed-sign behaviour of the 
known exact coefficients in the $\overline{\mbox{MS}}$ scheme.

The knowledge of the nature of the UV singularity, which we provide
in this note, could be used to optimise conformal mapping techniques 
to dispose of UV renormalon growth (for an application of 
conformal mappings 
in this context, see \cite{ALT95,SOP95}) or to improve Pade-type
approximations for the Borel transform of the perturbative series
by combining the information on singularities with the known low-order
coefficients.  The utility of this procedure may be limited 
since the leading UV asymptotic behaviour considered above is not
relevant for orders as low as $n=2$, the highest order known exactly.
This is especially so, because the 
leading behaviour is related to four-fermion 
operators. Subgraphs that contribute to the coefficients of these 
operators appear first at order $\alpha_s^2$. It takes several 
more orders to see the exponentiated effect of two-loop running 
of the coupling and the eigenvalues of the anomalous dimension 
matrix embodied in the factor $n^{\beta_1/\beta_0^2+\lambda_1}$.\\

{\bf Acknowledgements.}
We thank David Summers for reading the 
manuscript. The work by  N.~K. was supported by NORDITA as a part of the 
Baltic Fellowship program funded by the Nordic Council of 
Ministers.

\end{document}